\documentclass[a4paper,twoside]{article}
\pdfoutput=1
\usepackage{epsfig}
\usepackage{subcaption}
\usepackage{calc}
\usepackage{amssymb}
\usepackage{amstext}
\usepackage{amsmath}
\usepackage{amsthm}
\usepackage{multicol}
\usepackage{pslatex}
\usepackage{apalike}
\usepackage{url}
\usepackage{comment}
\usepackage{tabularx}
\usepackage[moderate]{savetrees}
\usepackage{xcolor}
\usepackage{SCITEPRESS}     % Please add other packages that you may need BEFORE the SCITEPRESS.sty package.

\begin{document}

\title{An Overview of Cryptographic Accumulators}

\author{\authorname{Ilker Ozcelik\sup{1, 2}\orcidAuthor{0000-0002-2032-1640}, Sai Medury\sup{1}\orcidAuthor{0000-0002-2349-8783} and Justin Broaddus\sup{1}\orcidAuthor{0000-0001-8540-997X}, Anthony Skjellum\sup{1}\orcidAuthor{0000-0001-5252-6600}} \affiliation{\sup{1}SimCenter \& Department of Computer Science and Engineering, University of Tennessee at Chattanooga, Chattanooga, Tennessee, USA}
\affiliation{\sup{2} Recep Tayyip Erdogan, Rize, Turkey}
\email{ozcelikilker@ieee.org, Sai-Medury@mocs.utc.edu,  \{Justin-Broaddus, Tony-Skjellum\}@utc.edu}
}

%AS-reviewed/edit 10/29/20 @ 12:41est
\abstract{
This paper is 
a primer on cryptographic accumulators and how to apply them practically.  
A cryptographic accumulator is a  space- and time-efficient data structure used for set-membership tests.  Since it is possible to represent any computational problem where the answer is yes or no as a set-membership problem, cryptographic accumulators are invaluable data structures in computer science and engineering. But, to the best of our knowledge, there is  neither a concise survey comparing and contrasting 
various types of accumulators nor a guide for how to apply the most appropriate one for a given  application. Therefore, we address that gap by describing cryptographic accumulators while presenting their fundamental and so-called optional properties. We discuss the effects of each property on the given accumulator’s performance in terms of space and time complexity, as well as communication overhead.
}


\onecolumn \maketitle \normalsize \setcounter{footnote}{0} \vfill

\section{\uppercase{Introduction}}
There are many use cases where one might need to maintain a list of elements for the purpose of determining whether an element being presented is part of this list or not. A common example is a list of credentials that have been authorized and granted certain privileges like an Access-Control List (ACL). During Authentication, an account management system will check to see whether the credentials entered are a part of the ACL or not and grant/deny privileges accordingly. If this list were small, with only a few hundreds of elements at any given time, it would not take long to load the entire list into memory,  compare each credential, and search for a match. The time complexity of this algorithm scales linearly ($O(n)$) with the size ($n$) of the list. Therefore, it will perform poorly if the list grows to a number in the hundreds of thousands (and the performance is controlled by I/O speed, not main memory speed if the list is sufficiently large). We can reduce this complexity to sublinear ($O(\log n)$) by doing certain pre-computations on the list, such as by ordering it, then performing a binary search. Sublinear time complexity is an acceptable computational complexity by ``industry standards'' but there is substantial overhead of sorting the elements that has the average computational complexity of $O(n\log n)$, which increases the total computational complexity of the algorithm to $O(n\log n)$. 

We can further reduce computation complexity by trading off memory space by constructing auxiliary data structures like hashmaps 
with constant time lookup complexity. This could be a great alternative to pre-computation overhead and facilitates constant time lookup, which is literally the best possible speed up. However, this approach comes with an overhead of having to store extra data in  memory that will also scale linearly ($O(n)$) with the size of the list.
Depending on the memory size of the processing unit, a SHA256 hashmap representing a list that contains, for instance, 10 million elements may not fit in the memory of a low-energy and resource constrained devices.

A cryptographic accumulator can be used as an alternative to search-based approaches. Cryptographic accumulators are space-efficient data structures that rely on cryptographic primitives to achieve sublinear time complexity 
for set-membership operations. They were first proposed to solve document time-stamping and membership-testing purposes  \cite{benaloh1993one}
Later, they were 
employed to implement authenticated data structures \cite{ghosh2014verifiable,goodrich2002efficient} privacy-preserving \cite{slamanig2012dynamic} and anonymity-conscious \cite{camenisch2002dynamic,miers2013zerocoin,sudarsono2011efficient} applications. Also, with the advent of blockchain technology, demand for data to be stored in a decentralized manner is rising rapidly and cryptographic accumulators are potentially well suited to support quick constant-time set-membership testing on such data.

Cryptographic accumulators use novel, probabilistic data structures for set-membership that minimize space complexity by compressing the hashmap, via a set of cryptographic functions, into a constant size bit-based data structures like the Bloom filter \cite{bloom1970space},
%\footnote{The Bloom filter is a space-efficient probabilistic data structure that can be applied to test if a member is part of a given set. It is a bit-array of size $m$ with all initialized to 0. And $k$-different hash functions define the unique combination of $k$ bits to set among the total $m$ array indices.},
or its more recently proposed alternative called the cuckoo filter \cite{fan2014cuckoo}. This compression is lossy, 
 inevitably giving rise to false positives from possible collisions.

Alternatively, one can use a cryptographic accumulator, a hash representative generated from the elements in the list, that is of constant size and provides constant-time lookup complexity without potential false positives \cite{benaloh1993one}. These types of cryptographic accumulators are better suited for distributed applications where a trusted authority is responsible for continuously maintaining the accumulator (hash representative); the trusted authority keeps the accumulator up to date with additions and deletions of elements from the list. Participants other than this authority could be clients themselves that are part of the list or else be verifiers that are trying to determine whether an element being presented is part of the list. Such cryptographic accumulators, also called asymmetric cryptographic accumulators \cite{kumar2014performances}, additionally require the generation of witnesses \cite{kumar2014performances} for each element. 

The witness is the value corresponding to an element that is required to verify an element's membership in the accumulator \cite{kumar2014performances}; it is unique to each element and needs to be updated each time an element is added or deleted to the accumulator. This witness can also be communicated to the client for storage and later presentation on-demand to a verifier. The trusted authority, often known as the Accumulator Manager (AM)  \cite{akkaya2018efficient},  needs to perform $O(n)$ operations to update the witnesses of $n$ elements for each new addition  (resp, deletion) of an element to (resp, from) the list; this is the trade off for having virtually constant size computation complexity and communication overhead between the verifier, client, and AM for determining set-membership.   
% REVISED INTRO TO HERE, 10/29/20{AS} @ 12:51est
%
Most  studies thus far regarding cryptographic accumulators have been  theoretical  and
consequently focus on their
underlying theories. By carefully combing through these studies and presenting a guide with a gentler learning curve, this paper provides an accessible discussion for those who are interested in learning, developing, and/or utilizing cryptographic accumulators in their applications.

The remainder of this paper is organized as follows: Section \ref{sec:crypto-accumulators} introduces the set membership problem and cryptographic accumulators. Different architectures of cryptographic accumulators and their classification are also presented in this section. 
Effects of ``optional features'' on accumulator operations are discussed in Section \ref{sec:cost-benefit-analysis}, in terms of memory usage, computational- and communicational-complexity. 
Finally, Section \ref{sec:conclusion} concludes with a discussion of current and potential applications.

For reference throughout, brief descriptions of the different kinds of accumulators can be found in Table~\ref{tab:kindsofaccumulators}.
Also, some notable terms used in this paper and their simplified definitions are listed in Table~\ref{tab:terms}, below.

\begin{table}[!h]
\renewcommand{\arraystretch}{1.1}
% \extrarowheight 
\caption{Kinds of
Cryptographic Accumulators}
\label{tab:kindsofaccumulators}
\centering
% Some packages, such as MDW tools, offer better commands for making tables
% than the plain LaTeX2e tabular which is used here.
\resizebox{\linewidth}{!}{%
\begin{tabular}{|l|p{1.5in}|}
\hline
\textbf{Accumulator} & \textbf{Key Properties} \\
\hline 
Strong  & One whose manager is considered not trusted by design. \\ 
& These  do not require trapdoors.
%for any of their operations or initializations. 
\\ \hline
Dynamic (resp, Static)  & One whose input set  can (resp, cannot) change in time \\ \hline
Additive (resp, Subtractive)  & One whose input set can only grow (resp, shrink) in time \\ \hline
%Subtractive  & One whose input set can only shrink in time \\ \hline
%Dynamic  & One whose input set can grow and shrink in time \\ \hline
Positive  & One that can only provide membership-proof  for accumulated elements \\ \hline
Negative  & One that can only provide non-membership-proof for non-accumulated elements \\
\hline
Universal & One that can provide membership (resp, non-membership) \\
& proofs for accumulated (resp,  non-accumulated) elements  \\ \hline
\end{tabular}}
\vspace*{-.2cm}
\end{table}
%\section{Basic Terminology}
%Some of the notable terms used in this paper and their simplified definitions are listed in Table~\ref{tab:terms}
\begin{table*}[!h]
\renewcommand{\arraystretch}{1.3}
% \extrarowheight 
\caption{Glossary 
of Cryptographic-Accumulator Terms used in this Paper}
\label{tab:terms}
\centering
% Some packages, such as MDW tools, offer better commands for making tables
% than the plain LaTeX2e tabular which is used here.
\resizebox{\linewidth}{!}{%
\begin{tabular}{|l|l|}
\hline
\textbf{Term} & \textbf{Definition} \\ \hline
Bezout coefficients & Coefficients a and b such that ax + by = d where d is the greatest common divisor of x,y \\ \hline
Collision & A situation where multiple different inputs to a hash function outputs the same hash value \\ \hline
False Positive (Type I Error) & In Set-Membership, incorrectly deciding the existence of an element even though it does not exist in the set \\ \hline
Filter & A signal processing approach used to remove unwanted components of input data \\ \hline
Hash function & A function  used to uniquely map a key to an index of an array that contains associating value \\ \hline
Hashmap (aka Hash Table) & 
%Also called a Hash Table, 
A data structure that associates keys to values 
\\ \hline 
Quasi-commutative property & A function has this property if it generates the same output for a set of same inputs regardless of their order. \\ \hline
Random Oracle & A theoretical black box that provides a truly random response (selected from an output domain) for every unique request\\ \hline
Trapdoor & In cryptography, special information needed to perform the inverse of a cryptographic operation efficiently \\ \hline
Trapdoor function & A one-way function that has a trapdoor \\ \hline
Witness & Auxiliary information   required to efficiently verify the authenticity of a statement or other unit of information 
 \\ \hline
\end{tabular}}
\end{table*}

\section{\uppercase{Cryptographic Accumulators}}\label{sec:crypto-accumulators}

As noted above, set membership tests (yes/no) tests) are used in many applications including
database, authentication, and validation systems. 
Testing set membership can be performed via a search  on the given set but this method can be a resource-intensive task as the set size increases. To address these limitations, researchers proposed cryptographic accumulators \cite{fan2014cuckoo,baldimtsi2017accumulators,tremel2013real,reyzin2016efficient}. 
The fundamental idea behind the accumulator is being able to accumulate values of a set $A$ into a small value $z$ in such a way that it is possible to prove only the elements of set $A$ have been accumulated \cite{fazio2002cryptographic}.

\subsection{Classification}
Cryptographic accumulators can be categorized as either symmetric or asymmetric. 
Symmetric Accumulators \cite{kumar2014performances} 
are designed using symmetric cryptographic primitives and can verify membership of elements without the need of a corresponding witness. The Bloom filter \cite{bloom1970space}---a type of array data structure---is    a symmetric cryptographic accumulator that uses $k$  hash functions that 
create a unique combination of indices in the array based on the input element; it provides a limited representation of set-membership with a false positive rate that grows as the number of elements in the list approaches the maximum capacity of the list \cite{bloom1970space}. Equation~\ref{eq:eq1} provides the estimate of the false positive rate of a simple Bloom filter construction:
\begin{equation}\label{eq:eq1}
    {\it FPR} = (1 - [1 - \frac{1}{m}]^{kn})^{k}\approx (1 - e^{\frac{-kn}{m}})^{k}
\end{equation}
with $m$ defined as the size of the array, $k$  the number of hash functions, and $n$  the number of accumulated elements \cite{bloom-filter-acm}. Variations of the Bloom filter \cite{fan2014cuckoo} have sought to minimize this false positive rate but are unable to eliminate it entirely.  Because Bloom filters are static accumulators,  they cannot accommodate growing list sizes, so they must be regenerated after reaching full capacity  during the transaction discovery process.
%\footnote{As an example, a Bloom filter is  employed in Bitcoin \cite{bitcoin-spv} to help Simplified Payment Verification (SPV) clients communicate with full nodesA full node in the Bitcoin network follows the safest security model by downloading and maintaining a copy of the entire blockchain from the Genesis block till the most recent block.}.

\begin{figure*}[ht]
\begin{center}
\includegraphics[width=0.7\textwidth]{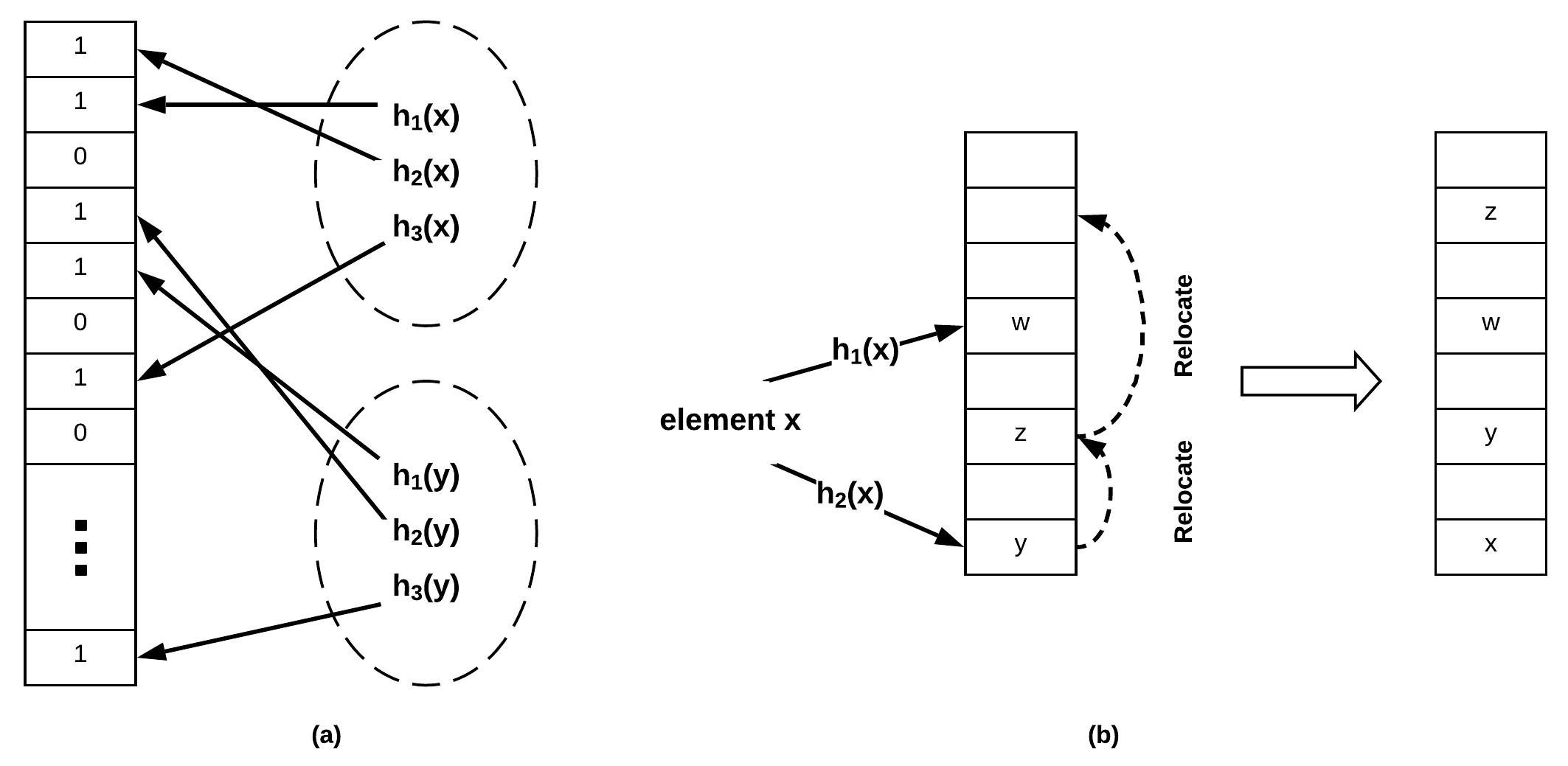}
\caption{Bloom and cuckoo filter accumulation processes. (a) Adding 2 elements into a Bloom filter using 3 hash functions. (b) A cuckoo filter before and after adding element x using two hash functions.}
\label{Fig:filters}
\end{center}
\end{figure*}

The recently proposed cuckoo filter \cite{fan2014cuckoo} is a dynamic data structure that functions similarly to the simple Bloom filter but with additional capabilities such as the ability to delete elements. A cuckoo filter implements a cuckoo hash table \cite{fan2014cuckoo} to save fingerprint representations of the elements in an accumulated set. A cuckoo hash table is an array of buckets in which each stored element is associated with two indices in the array. Two associated indices allow for the dynamic rearrangement of the elements stored in the cuckoo hash table, providing optimized space efficiency and low false-positive rates. For a given number of elements, a cuckoo filter outperforms a space-optimized Bloom filter in terms of false-positive rate and space overhead \cite{fan2014cuckoo}. Bloom and cuckoo filter accumulation processes are presented in Figure~\ref{Fig:filters}a and Figure~\ref{Fig:filters}b, respectively.

Asymmetric Accumulators \cite{kumar2014performances} require witness creation and update for dynamic verification of set membership \cite{baldimtsi2017accumulators}.  They are built on asymmetric cryptographic primitives \cite{baldimtsi2017accumulators} and require the underlying hash algorithm to exhibit the quasi-commutative property \cite{benaloh1993one}.   
The RSA accumulator \cite{benaloh1993one} is one example of an asymmetric accumulator that uses RSA modular exponentiation to achieve the quasi-commutative property. A simple RSA accumulator construction consists of the following expression for addition:
\begin{equation}
acc_{n}=acc_{n-1}^{x} \mod ~N,
\end{equation}
where $acc_{n}$ is the new accumulator value after addition, $acc_{n-1}$ is the old accumulator value before addition, $x$ is the element being added, and $N=pq$ where $p$ and $q$ are considered to be strong prime numbers whose Sophie Germain prime numbers $p' = \frac{p-1}{2}$ and $q' = \frac{q-1}{2}$ are the accumulator trapdoor. One drawback of an RSA accumulator is that it is collision-free only when accumulating prime numbers. A prime representative generator is required to accumulate composite numbers without collision
%\footnote{The bilinear-map accumulator introduced by Nyugen in \cite{nguyen2005accumulators} uses bilinear pairings to achieve the quasi-commutative property. It can accumulate composite numbers without the need to generate prime representatives. We should note that the majority of the asymmetric accumulator constructions provided by the literature have been based on RSA and Merkle Trees.}
. 
The implementation of an RSA accumulator by \cite{tremel2013real} a random oracle prime representative generator provided by 
\cite{baric1997collision} was used. Asymmetric accumulators can be further classified based on operations supported and the type of membership proofs provided. This classification will be further explained and analyzed in the following sections.

A Merkle hash tree can also be implemented as an asymmetric accumulator to prove set membership of elements \cite{baldimtsi2017accumulators}. It is classified as an asymmetric accumulator because a member of its set requires a witness to prove membership (or non-membership). But, it does not use asymmetric cryptographic primitives nor does it need the underlying hash function to exhibit the quasi-commutative property. 
The root node of the Merkle Tree is called the Merkle root and its value is the pairwise accumulated hash of all of the non-root nodes in the tree. The Merkle root value must be recalculated when there is an addition/deletion of a member in the set \cite{merkle1989certified}. Checking for set membership can be done with a portion of the tree \cite{merkle1989certified}, making it unnecessary to download the full data structure.
\subsection{Architectures}
There are two well known cryptographic accumulator architectures: the one-way  and the collision-free accumulator. Both are discussed in turn.
\subsubsection{One-way Accumulator}
\cite{benaloh1993one} proposed the first-ever accumulator construction, known as the \emph{One-way Accumulator},   characterized as a family of one-way hash functions with the additional quasi-commutative property. 
A one-way hash function $H$ is a function that can accept an arbitrarily large message $M$ and returns a constant size output that is also called a message digest $MD$. For   H to be a one-way hash, it must satisfy the following properties \cite{onewayFunc,onewayhashFunc}:
\begin{itemize}
    \item The description of function $H$ should be public and should work without needing to know secret information. 
    \item For a message $M$, it must be easy to calculate its message digest, $MD = H(M)$.
    \item Given an $MD$, it must be difficult to determine $M$ for a range where $H$ is valid.
    \item For any $M$, the probability of finding an ${M}^\prime\neq M$ such that ${H}^\prime(M) = H(M)$ is negligible. 
\end{itemize}
The quasi-commutative property is a generalization of the commutative property. A function $h$ defined as
%\vspace{-.2cm}
%\begin{equation*}
   $h: X \star Y \Rightarrow X$
%\end{equation*}
holds the quasi-commutative \cite{benaloh1993one} if  $\forall x \in X $ and  $\forall y_1, y_2 \in Y$,
%\vspace{-.2cm}
\begin{equation}
    h(h(x,y_1),y_2) = h(h(x,y_2),y_1).
\end{equation}
This property ensures cryptographic accumulator to generate same digest even if set members are accumulated in different orders.
Despite using a strong one-way hash function \cite{fazio2002cryptographic}, Benaloh and de Mare's accumulator %is 
can be compromised if 
an attacker can choose a subset of the values being accumulated \cite{fazio2002cryptographic}. 
\cite{baric1997collision} proposed “Collision-free accumulators” to address this issue, described next.

\subsubsection{Collision-free Accumulator}
 One-Way accumulators are elementary and ideal for set membership, whereas collision-free accumulators are more general \cite{baric1997collision}, and so are better suited for designing fail-stop signature schemes\footnote{Fail-Stop Signatures \cite{baric1997collision}, as opposed to conventional digital signature schemes, allow a signer to produce proof-of-forgery when a signature has been forged by using massive computational power to find collisions. The proof further shows that the computational assumption no longer stands (fails) and the signature scheme that was used can be stopped from being valid in future communication.} 
and Group Signatures. Collision-free accumulators consist of an accumulator scheme that is defined together by four polynomial-time algorithms: Gen, Eval, Wit and Ver~\cite{fazio2002cryptographic}. We discuss these in turn.

The key generation algorithm \emph{Gen}~\cite{fazio2002cryptographic} is used to generate the necessary key for a desired size accumulator, which accepts a security parameter ($1^{\lambda}$) and accumulator threshold value $N$. The threshold value defines the maximum number of elements that can be accumulated securely. Gen returns a key $k$ from key space $K_{\lambda N}$.
The evaluation algorithm $Eval$~\cite{fazio2002cryptographic} is used to accumulate elements of set $A = {y_{1}, \cdots , y_{N'}}$ where $N'\leq N$. It accepts the accumulator key $k$ and values to be accumulated, ${y_{1}, \cdots , y_{N'}}$, as input. Eval returns the accumulated value $z$ and an auxiliary information, $aux$ that will be used by other algorithms. It is important to note that the Eval algorithm must return the same $z$ value for the same input, but it may generate different auxiliary information. 
The witness extraction algorithm $Wit$~\cite{fazio2002cryptographic} generates the witness for a given input. It accepts the accumulator key $k$, the input value $y_{i} \in Y_{k}$ and the previous auxiliary and accumulator value $z$ generated by Eval. Wit returns a witness value $w_{i} \in W_{k}$ to show input $y_{i}$ was accumulated. Otherwise it returns the symbol $\perp$.
The verification algorithm $Ver$ is used to test the existence of an element in an accumulator; it accepts the accumulator key $k$, accumulator $z$, input value $y_{i}$, and its corresponding witness $w_{i}$. Ver returns the value TRUE or FALSE. 

The accumulator scheme is paired with the property of \emph{collision freeness}~\cite{fazio2002cryptographic}. Collision freeness ensures that an adversary bounded by polynomial-time cannot generate a set of values  $Y = {y_{1}, y_{2}, \cdots , y_{N} }$ that produces an accumulator value $z$ that allows for a value $y \notin Y$ and a witness $w$ that allows for $y$ to be proven as a member accumulated in $z$.

\cite{baric1997collision} propose two theoretical constructions of collision-free accumulators for building Fail-Stop Signature schemes. There also exists a dynamic accumulator implementation by %Camenish and Lysyanskaya 
\cite{camenisch2002dynamic} that was inspired by the Collision-Free accumulator architecture but was meant for set-membership testing.
\begin{comment}
\begin{figure*}
    \centering
    \includegraphics[width=0.9\textwidth]{Figs/CryptographicAccumulatorsOverview.pdf}
    \caption{An overview of actors, variables, and operations involved in cryptographic accumulators}
    \label{fig:cryptoacc-overview}
\end{figure*}
\end{comment}
\subsection{\uppercase{Properties}}
Cryptographic accumulator properties are discussed next. First security properties are considered., then ``optional properties.'' 
\subsubsection{Security Properties }
There are four prominent security properties  of asymmetric cryptographic accumulators: \emph{soundness, completeness, undeniability and indistinguishability}~\cite{derler2015revisiting}.

The \emph{Soundness} (aka Collision-Freeness)
is defined as the probability of computing a membership witness for an element that is not part of the accumulated set or non-membership witness for an element part of the accumulated set is negligible \cite{derler2015revisiting}. 

The \emph{Completeness}, (aka Correctness), property requires that all honestly accumulated values  be verified as true with their respective witnesses with a negligible probability of error \cite{derler2015revisiting}. 
An accumulator is called \emph{undeniable} if the probability of computing a membership and non-membership witnesses together, of the same input, is negligible. Undeniability implies the collision-freeness property but not all collision-free accumulators have the undeniability property \cite{derler2015revisiting}. 

\emph{Indistinguishability} is both a security- and privacy-related property. An accumulator is indistinguishable if no information about the accumulated set is leaked by either the accumulator or its witnesses. This can be achieved by either accumulating a random value from the accumulation domain or using a non-deterministic Eval Algorithm \cite{derler2015revisiting}.  

\subsubsection{Optional Crypto-Accumulator Features}\label{subsec:optional}
\begin{figure*}[ht]
\begin{center}
\includegraphics[width=0.9\textwidth]{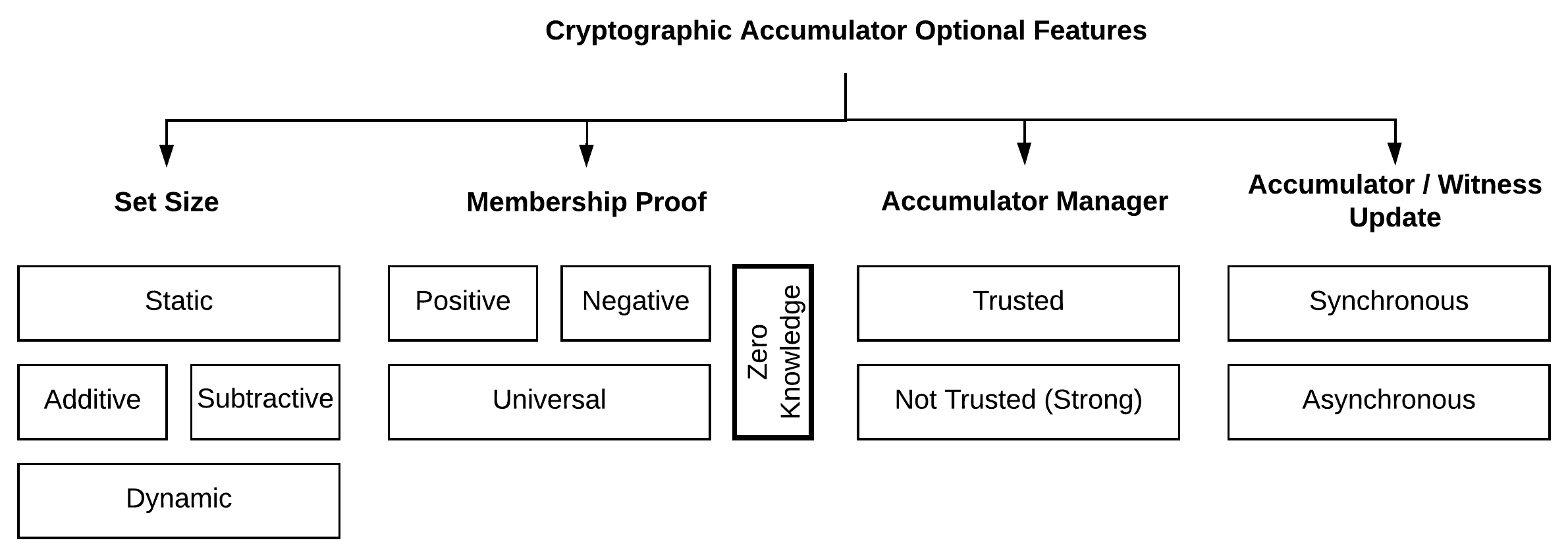}
\vspace*{-.1cm}
\caption{Cryptographic Accumulator Optional Features}
\label{Fig:optionalf}
\end{center}
\vspace*{-.35cm}
\end{figure*}

 Constructions can be optimized 
by
selecting a subset of available features.
Each available feature comes with a cost that can significantly impact the system design and implementation.
   Notable  features are presented in Figure~\ref{Fig:optionalf} and discussed next. 

The set of accumulated elements 
is not always static; 
new elements may be added to or removed from this set over time. This is a common case for applications requiring to grant / revoke the privilege of a credential
If an accumulator only supports additions, it is termed \emph{additive}. Similarly, an accumulator that only supports deletions is termed  \emph{subtractive}. Addition and subtraction can be performed by redoing the accumulation process after updating the set. But, this approach is generally not practical, since recalculation %will take 
takes polynomial time and depends on the size of the accumulated set \cite{baldimtsi2017accumulators}. Dynamic accumulators are those that can efficiently (in sublinear or constant time complexity) update the accumulator and the respective witness values when a new element is added to (resp, removed from) the accumulated set \cite{au2009dynamic}.  Accumulators that do not support additions or deletions are termed \emph{static}.  

Accumulators are alo termed \emph{positive}, \emph{negative}, or \emph{universal} based on the type of membership proof(s) they can support \cite{baldimtsi2017accumulators}. Positive accumulators  only support set membership proof of inclusion. Thus, for all elements in the accumulated set, there exists an efficiently computable witness $w$. Negative accumulators can only support a non-membership 
proof.  For all elements that are considered non-members in regards to the accumulated set, there exists an efficiently computable witness $w^\prime$. Lastly, a universal accumulator supports both membership and non-membership proofs \cite{li2007universal}. 

The Zero Knowledge Proof (ZKP) is a privacy preserving membership proof used by cryptographic accumulators. It was initially proposed by  
 \cite{goldwasser1989knowledge} in 1989. With ZKP, it is possible to show the accuracy of a statement about a secret without revealing the secret itself. 
This is possible because, if one can compute the same output that the prover provides by only accessing the input of the verifier; it should also be possible to compute the output before such interaction occurs. Therefore, through ZKP systems, an honest verifier need not interact with the prover \cite{morais2018zero} 
 to verify the accuracy of a statement. 

A single accumulator data structure alone will be insufficient to implement ZKPs but two or more cryptographic accumulator schemes can be used to implement an ZKP system. Even with a combination of accumulators, the interaction mechanisms must be augmented to carry out ZKPs. Additionally, these interactions must be standardized to make all parties aware of these mechanisms.

The zero-knowledge property has been proven to imply indistinguishability for cryptographic accumulators \cite{ghosh2016zero}. But, maintaining the integrity, zero knowledge property, and the efficiency of the accumulator simultaneously is  challenging.  This complexity can be reduced by restricting the accumulator’s design to a trusted setup \cite{ghosh2016zero}, %. However, with trusted setup,
where the accumulator value is always maintained by a trusted party. %thus the reliability of the accumulator is not a concern.
But, a trusted user is required to generate a secret value called a \emph{trapdoor} to  compute the accumulator and witness values efficiently. This  setup  raises  major security and centralization concerns. But, there is no practical trapdoorless (strong) accumulator that can produce constant-size proofs. 
In trapdoorless accumulators, since the accumulator manager is also considered untrusted,
the witness size grows at least like $\log N$ where $N$ is the number of accumulated elements  \cite{ghosh2016zero}.

Applications can be either local, where only a single entity/authority is responsible for proving membership, or distributed 
where multiple parties in a network interact with each other. In %a 
the distributed case, the communication channel between all relevant parties 
becomes a bottleneck because membership witness additions or updates 
must be broadcast 
each time an element is added to or deleted from the accumulator. To 
minimize this communication overhead, asynchronous accumulators  were proposed for distributed applications \cite{reyzin2016efficient}.

The asynchronous accumulator  relies on low update frequency and 
compatibility with old accumulator version \cite{reyzin2016efficient}.  
An accumulator has a low update frequency property, if the witness of an element requires less  updates compared to the number of elements added after the element \cite{reyzin2016efficient}.
% Added by Sai on Dec 16 2020
Low update frequency can be achieved only if it is additionally possible to accurately verify membership using an outdated accumulator value, in other words, the accumulator must be backwards compatible.
%\footnote{An outdated accumulator value is an accumulator value that does not reflect the most recently updated accumulated set. It is the resulting value of a previous update to the accumulator.}
It is important to note that the element tested for membership proof must be accumulated before the outdated accumulator value was generated \cite{reyzin2016efficient}.
\section{\uppercase{Cost-Benefit Analysis}}\label{sec:cost-benefit-analysis}

Security properties of cryptographic accumulators control the reliability of the accumulator design. It is expected for an accumulator to satisfy these properties. However, some of the features presented in Section~\ref{subsec:optional} are design choices rather than necessities.  
We encountered four major design choices in the literature,
and each choice affects the
implementation complexity and overall accumulator performance. 
In this section, we present   and discuss effects of these design choices in terms of space complexity, time complexity, and communication overhead. 

In an application, the input set that needs to be accumulated can be static, additive, subtractive, or dynamic. A static set is a set whose list of elements do not change through time. In additive and subtractive sets, elements can only be added to or deleted from the set respectively. A dynamic set refers to the ability  both to add and delete elements from the set. 
Since a static set does not cause system space, time complexity and communication overhead, we will not consider it in our discussion.

\begin{table}[!t]
\renewcommand{\arraystretch}{1.3}
% \extrarowheight 
\caption{The space- and time-complexity and communication overhead in accumulators with input set changes. Independent cases are separated using vertical line ($\mid$) and mutually exclusive cases are separated using virgules (/).}
\label{tab:cb-inputSet}
\centering
% Some packages, such as MDW tools, offer better commands for making tables
% than the plain LaTeX2e tabular which is used here.
%\resizebox{0.75\linewidth}{!}{%
\resizebox{\linewidth}{!}{%
\begin{tabular}{|l|l|l|l|}
\hline
 & \textbf{Additive} & \textbf{Subtractive} & \textbf{Dynamic} \\ \hline
\begin{tabular}[c]{@{}l@{}}Space Complexity \\ ($Size_{Acc} \mid Size_{Wit} \mid Size_{AM}$)\end{tabular} & $O(1) \mid O(1) \mid O(1) $ & \multicolumn{1}{c|}{$ O(1) \mid O(1) \mid O(N) $} & $ O(1) \mid O(1) \mid O(N) $\\ \hline
\begin{tabular}[c]{@{}l@{}}Time Complexity at AM\\ ($Wit_{Member} \mid Wit_{Non-member}$)\end{tabular} & $ O(1) \mid O(N) $ & $ O(N) \mid O(N) $& $O(1) \mid O(N) \mid O(N)$ \\ \hline
\begin{tabular}[c]{@{}l@{}}Communication Overhead \\ At AM\end{tabular} & $ O(1)$ & $ O(N) $ & $ O(1) | O(N) $ \\ \hline
\end{tabular}
}
\end{table}

The effects of a dynamic input set over system space, time complexity, and communication overhead is presented in Table~\ref{tab:cb-inputSet}. This table is compiled considering major asymmetric cryptographic accumulators with the exception of the tree-based strong accumulator, the Merkle Tree. More information about Merkle Tree based accumulators can be found later in this section, in the discussion about the accumulator manager (AM) trust (see Table \ref{tab:cb-trust}).
\begin{comment}
~{\color{blue} We do not have separate subsections for discussions. Can we contınue the sentence like} 
{\color{green} in Table~\ref{tab:cb-trust}.}  {\color{red} WHERE DO I FIND THIS? DO YOU MEAN TO REFERENCE SOME OTHER PART OF THE PAPER? If so, please do so.}.
\end{comment}

When we evaluate space complexity based on a dynamic input set, we will investigate accumulator, witness and accumulator manager storage size changes (see Table~\ref{tab:cb-inputSet}). The size of an asymmetric accumulator can be specified using an accumulator threshold value $N$ during accumulator generation. This value defines the upper bound on the total number of values that can be accumulated securely in the accumulator \cite{fazio2002cryptographic}. Adding and removing elements to/from these accumulators do not impact the accumulator and witness size. On the other hand, RSA- and Bilinear-Map-based accumulators require a trapdoor upon  initialization of the accumulators themselves as well as during the deletion of a member in the accumulated set. 
In otherwise trapdoorless constructions, an
accumulator manager's storage can grow proportional to the set size.

For every element added to the set, the accumulator manager must update the accumulator value and create a membership witness. The accumulator manager must also update the accumulator value after each deletion of an element and provide a non-membership witness if supported by the accumulator construction. Accumulator update upon element addition and subtraction and generating a membership witness for a new element are constant time processes performed by the accumulator manager. On the other hand, generating a non-membership witness for RSA and Bilinear-Map accumulators  requires  time proportional to the number of elements in the set. 

Asymmetric accumulators provide a constant verification and witness update time either for the addition or subtraction of an element. However, the witness-update process may cause a bottleneck in RSA and Bilinear-Map accumulators during element subtraction since the witnesses have to be updated by the accumulator manager itself. However, there are proposed constructions in the literature that allow for the accumulated members to update their own witnesses after deletion by using Bezout coefficients \cite{baldimtsi2017accumulators,li2007universal}. 

In synchronous accumulators, the AMs must broadcast 
information about new elements to witness holders. 
This communication overhead 
is minimized in asynchronous accumulators, since witness holders 
need not  
update accumulator/witness value as frequently. 
But, RSA and Bilinear-Map accumulators experience additional communication complexity after element subtraction because of the queries from witness holders needed to retrieve updated witnesses from the accumulator manager.

Cryptographic accumulators can be  positive, negative, or universal based on the membership proof they provide.
Positive Accumulators provide only membership proofs to determine whether an element has been accumulated to the set of members. Negative accumulators, on the other hand, provide only non-membership proofs to determine whether an element has been accumulated to the set of non-members. Universal accumulators can provide both membership and non-membership proofs. Table~\ref{tab:cb-membership} describes the  time complexity,  space complexity, and communication overhead of an accumulator based on the (non-)membership proof. 
Table~\ref{tab:cb-membership} shows the minimum and maximum values observed across all membership proof types. 
To identify generalized operations found across a class of accumulators with similar applications, we limited the scope by selecting dynamic accumulators that require a trusted setup with an accumulator manager (AM) and synchronized communication. 
Also, the values for the Merkle Tree accumulator were omitted from Table~\ref{tab:cb-membership}; they will be discussed, however, in the evaluation of Table~\ref{tab:cb-trust}.
\begin{table}[!t]

\renewcommand{\arraystretch}{1.3}
% \extrarowheight 
\caption{The space-, time-complexity and communication overhead experienced by accumulator managers (AM) with different membership proof types.
%The effects of different membership proof types on space-, time-complexity and communication overhead of an accumulator.
%Here, we use the following denotations: $ a $ - number of elements added; $ d $ - number of elements deleted; $ S $ -  total number of member elements (Note that $ S =  a  -  d $)
Symbols $a$, $d$, and $S$ $(a - d)$ denote the number of elements added, deleted, and present  respectively. ‘$\sim$’ represents the range %with minimum value on the left and maximum value on the right.
}
\label{tab:cb-membership}
\centering
% Some packages, such as MDW tools, offer better commands for making tables
% than the plain LaTeX2e tabular which is used here.

\resizebox{\linewidth}{!}{%
%\resizebox{0.6\linewidth}{!}{%
\begin{tabular}{|l|l|l|l|}
\hline
\textbf{AM Complexity/Work} & \textbf{Positive} & \textbf{Negative} & \textbf{Universal} \\ \hline
Space Complexity  & $ O(1) \sim O(d) $ & \multicolumn{1}{c|}{$ O(S) $} & $ O(S) \sim O( a ) $\\ \hline
Time Complexity & $ O(1) $ & $ O(1) $ & $ O(1) \sim O(S) $ \\ \hline
\begin{tabular}[c]{@{}l@{}}Communication Overhead \\ \end{tabular} & $ O(d) \sim O(a+d) $ & $ O(a  +  d )$ & $ O(a  +  d) $ \\ \hline
\end{tabular}
}
\vspace*{-.1cm}

\end{table}

The values in Table~\ref{tab:cb-membership} convey the range of  performance, including bottlenecks, one may achieve by utilizing positive, negative, or universal accumulators. The Universal Accumulator has the highest complexity in all three dimensions when compared with (Strictly) Positive or (Strictly) Negative accumulators. 
However, the time complexity in Universal Accumulators may be reduced by implementing a separate accumulator of positive and negative types.

The space complexity  for applying positive cryptographic accumulators is of the order of $O(1)$ for positive accumulators with the exception of BraavosB's \cite{baldimtsi2017accumulators} construction, which results from  its reliance on the Range-RSA
%\footnote{The Range-RSA accumulator represents the valid member ranges between the deleted members of the accompanying positive accumulator in BraavosB \cite{baldimtsi2017accumulators}. The accumulator manager must store the membership witnesses to the Range-RSA accumulator in order to support the necessary operations, thus creating a space complexity of $O( d  )$.}
construction for accumulating deleted elements. For negative and universal accumulators, space complexity is generally $O(S)$, which is the number of elements that are part of the accumulator at any given moment (not including elements that were deleted).% after being added to the accumulator). 

The computation time complexities mentioned in Table~\ref{tab:cb-membership} show that all operations are typically of  $O(1)$
except for the NonMemWitCreate operation, which is the most expensive operation if using Universal accumulators built upon RSA and Bilinear-Mapping cryptographic algorithms. 

The communication overhead is $O(a+d)$ for dynamic accumulators, regardless of whether they are positive or negative or universal, which results from the requirement for communication   for Member Witness updates after both additions and deletions of elements. For positive accumulators like CL-RSA-B, Braavos or BraavosB that are custom designed to minimize communication, the communication overhead is $O(d)$ because they do not require communication after addition of elements but only after deletion of elements. 

Based on system requirements, a cryptographic accumulator can be designed either on the foundation of a trusted or untrusted accumulator manager. The RSA and Bilinear-Map implementations are restricted to only trusted accumulator manager setups as a result of their reliance on a  secret trapdoor value for initialization and constant time operations. The only known family of accumulators that can be safely used in an untrusted system are those based on Merkle Trees. This is because Merkle Trees do not require trapdoor values at any point during their operations. 

\begin{table}[!t]
\renewcommand{\arraystretch}{1.3}
% \extrarowheight 
\caption{The space-, time-complexity and communication overhead in accumulators with Trusted AM and Untrusted AM.
%Effects of AM trust choice to space and time complexity and communication overhead of an accumulator. Here, $ a $ denotes the number of elements added to the accumulator; $ d $ denotes the number of elements deleted from the accumulator; while $ S $ denotes the total number of member elements in the accumulator (Note that $ S =  a  -  d $).}
Symbols $a$, $d$, and $S$ $(a - d)$ denote the number of elements added, deleted, and present  respectively. %‘$\sim$’ represents the range
}
\label{tab:cb-trust}
\centering
% Some packages, such as MDW tools, offer better commands for making tables
% than the plain LaTeX2e tabular which is used here.
%\resizebox{0.6\linewidth}{!}{%
\resizebox{\linewidth}{!}{%
\begin{tabular}{|l|l|l|}
\hline
 & \textbf{Trusted AM} & \textbf{Untrusted (Strong) AM} \\ \hline
\begin{tabular}[c]{@{}l@{}}Space Complexity \\ ($Size_{Acc} | Size_{Wit} | Size_{AM}$)\end{tabular} & $ O(1) \mid O(1) \mid O(S) $ & \multicolumn{1}{c|}{ $ O(1) \mid O(\log a ) \mid O( a ) $} \\ \hline
\begin{tabular}[c]{@{}l@{}}Time Complexity\\ (For all operations)\end{tabular} & $ O(1) $ & $ O(\log a ) $\\ \hline
Communication Overhead & $ O( a  + d ) $ & $ O(( a  +  d ) \log a ) $ \\ \hline
\end{tabular}}
\vspace*{-0.4cm}
\end{table}

Trusted accumulator manager systems use accumulators that do not provide the strong accumulator property. Accumulators based on RSA \cite{baricRSAAccumulator} and Bilinear Maps \cite{bilinearPairingBasedAccumulator} fall into these types of accumulators. As shown in Table~\ref{tab:cb-trust}, trusted systems are relatively more scalable with 
$O(1)$ time and space complexities.
And, in similarity to systems without trusted AM, the communication overhead 
grows linearly with the number of member additions and deletions.
However, systems with trusted accumulator managers (AM) are centralized.
 AM is required to maintain the integrity of the accumulator by 
safeguarding the trapdoor value \cite{camenisch2002dynamic}. 
In case the trusted AM is compromised, then the entire system may be brought down. Acquisition of the trapdoor value would allow an adversary to produce membership witnesses that are compatible with the accumulator value even for non-accumulated elements, threatening the \emph{soundness} \cite{derler2015revisiting} of the accumulator. 
This centralized infrastructure also limits the system’s ability to distribute computational load, placing more responsibility 
on the trusted accumulator manager.

Untrusted accumulator manager (AM) systems use strong (trapdoorless) \cite{derler2015revisiting} accumulators like Merkle Trees \cite{merkle1989certified}. 
Their operations require $O(\log a )$ time complexity;
they produce witnesses of size $O(\log  a )$, and 
their communication overhead is of superlinear time complexity of $O(( a  +  d )\log a )$.  
Systems with untrusted AMs require relatively more space and can be 
expensive to maintain. 
The size of a Merkle Tree scales like $O( a  )$ 
and does not reduce in size after deletions. 
 Thus, the Merkle Tree will continue to grow regardless of the number of deletions. 

Advanced security can be considered as the the primary advantage of untrusted accumulator manager systems. 
The use of a trapdoor is not required in a strong accumulator \cite{derler2015revisiting}, and the responsibility of maintaining the integrity of the accumulated set is distributed amongst all participating untrusted accumulator managers. 
 Further, the absence of a trapdoor  allows for distribution of the computational load 
among the untrusted managers and accumulated members. 

\begin{table*}[!t]
\renewcommand{\arraystretch}{1.3}
% \extrarowheight 
\caption{The space-, time-complexity and communication overhead in accumulators with varying frequency in %accumulator %and witness
updates.
Symbols $a$, $d$, and $S$ $(a - d)$ denote the number of elements added, deleted, and present  respectively. ‘$\sim$’ represents the range
%Effects of accumulator and witness update frequency to space,time complexity and communication overhead of an accumulator. $ a $ denotes the number of accumulated elements. $ d $ - denotes the number of deleted elements. $ S $ denotes the actual number of elements represented by the accumulator (Note that $ S $ is $ a $ - $ d $). ‘$\sim$’ represents the range with minimum value on the left and maximum value on the right.}
}
\label{tab:cb-update}
\centering
% Some packages, such as MDW tools, offer better commands for making tables
% than the plain LaTeX2e tabular which is used here.
\resizebox{0.8\linewidth}{!}{%
%\resizebox{\linewidth}{!}{%
\begin{tabular}{|l|l|l|l|}
\hline
 & \textbf{\begin{tabular}[c]{@{}l@{}}Partially \\ Asynchronous \\ (Trusted)\end{tabular}} & \textbf{\begin{tabular}[c]{@{}l@{}}Asynchronous \\ (Untrusted)\end{tabular}} & \textbf{Synchronous (Trusted \& Untrusted)} \\ \hline
\begin{tabular}[c]{@{}l@{}}Space Complexity \\ ($Size_{Acc} | Size_{Wit} | Size_{AM}$)\end{tabular} & $ O(1) \mid O(1) \mid O(1) $ & \multicolumn{1}{c|}{ $ O(\log a ) \mid O(\log a ) \mid O( a ) $} & $ O(1) \mid O(1) \sim O(\log a ) \mid O(S) \sim O( a ) $ \\ \hline
\begin{tabular}[c]{@{}l@{}}Time Complexity\\ (For all operations)\end{tabular} & $ O(1) $ & $ O(\log a ) $ & $ O(1) \sim \log a $ \\ \hline
Communication Overhead & $ O( d ) $ & $ O(\log a  + \log d ) $ & $ O( a  +  d ) \sim O(( a  +  d ) \log  a ) $ \\ \hline
\end{tabular}
}
\vspace*{-.3cm}
\end{table*}

An asynchronous accumulator must hold both the low-update-frequency property and 
backwards compatibility property 
\cite{reyzin2016efficient}. To the best of our knowledge, the only construction that provides a fully asynchronous accumulator was defined by \cite{reyzin2016efficient}. Another asymmetric accumulator that exhibits the low-update-frequency property but not the old accumulator compatibility property is the CL-RSA-B accumulator, presented by  \cite{camenisch2002dynamic}. We refer to accumulators that only hold the low-update-frequency property  as \emph{partially asynchronous}.  
These accumulators are notably only positive accumulators 
\cite{camenisch2002dynamic}. But, we are not claiming that the construction of a negative or universal asynchronous accumulator is infeasible; this remains an open question. 

The asynchronous accumulator defined by 
 \cite{reyzin2016efficient} 
takes the form of a dynamic set of Merkle Trees. The number of Merkle Trees grows by a factor of $D = log_{2}(n + 1)$. Each subsequent Merkle Tree holds varying numbers of members and the entire construction follows a combinatorial approach when additions are made \cite{reyzin2016efficient}. The benefit of using this accumulator is the reduced communication overhead by a factor of $O(\log a + \log b) = O(\log ab)$, 
%+ \log b )$ 
as shown in Table~\ref{tab:cb-update}. The asynchronous accumulator 
can also
 be applied in decentralized, untrusted systems because of its strong accumulator construction. The disadvantages are the dynamic sizes of the accumulator value and witnesses by a factor of $O(\log a )$. The accumulator manager (AM) also requires storage of $O( a )$. Verification takes $O(\log a )$ as well. To support deletion, a list of all member values must also be stored and maintained \cite{reyzin2016efficient}.

The partially asynchronous variant in the form of the CL-RSA-B accumulator has the advantage of constant time operations and constant size accumulator and witness values. It also only has to update witnesses after a member is deleted from the set; witness updates are not required after additions. For an accumulator scheme that is not expecting a significant number of deletions compared to additions, this feature is ideal because the communication overhead is only $ d $ as shown in Table~\ref{tab:cb-update}. The main drawback of a partially asynchronous accumulator is its inability to exhibit the old accumulator compatibility property, so the most up-to-date accumulator value must be held. Another disadvantage is its restriction to trusted accumulator schemes in which the trusted accumulator manager is the only entity that can generate witnesses upon the addition of new members and update the accumulator value when a member is deleted. This is due to the requirement of a trapdoor to perform deletions in a CL-RSA-B accumulator. Once a member is deleted, the new accumulator value as well as the deleted member’s ID is broadcasted to the other members to update their respective witness values \cite{baldimtsi2017accumulators}. 

In comparison with synchronous accumulators,  overall system's communication cost  is reduced in both asynchronous and partially asynchronous accumulators. Synchronous accumulators never have a communication overhead less than $O( a  +  b )$. This is because all witnesses must be updated after a member addition or deletion  in the set.
%and witnesses are incompatible with old {\color{red} 'old' means what?} accumulator values. 
The decision of which asynchronous accumulator type to pick comes down to whether the system requires a trusted or untrusted setup. Dynamic or static sizes of witness and accumulator values should be taken into consideration too.
\section{\uppercase{Known Implementations and Resources}}\label{sec:known-implementations-and-resources}

We provide a brief survey of known implementations and other resources here.

First, 
 Baldimtsi et al. proposed an accumulator called Braavos \cite{baldimtsi2017accumulators}; it is an RSA-based accumulator that is used in IBM’s Idemix anonymous credential system. The authors provided insight into the mechanisms of an RSA accumulator and its CL-RSA-B variant. They also presented a performance analysis of several asymmetric cryptographic accumulators and discussed the differences between them.
Second, 
Tremmel presented a quantitative comparison between the RSA and Bilinear Map accumulators \cite{tremel2013real}. He discussed the use of parallelization to aid in the computational cost of managing witness values with a centralized accumulator manager.  This work is complemented with a GitHub repository \cite{etremel-github-cryptoacc}. 
Third,
Fan et al.\@ discussed two symmetric accumulators, Cuckoo filter and Bloom filter, and proposed that Cuckoo filter is more effective in most use cases \cite{fan2014cuckoo}. These authors provided quantitative data to support their proposal as well as a GitHub repository \cite{cuckoo-filter-github} 
containing the source code of a Cuckoo filter implementation. 
Lastly, 
Reyzin et al.\@ introduced the concept of an asynchronous accumulator and defined the properties needed to achieve a sufficient construction \cite{reyzin2016efficient}.  They  also %authors 
discussed the drawbacks of synchronous accumulators in distributed public key infrastructure (PKI)
and outlined their detailed construction for a fully asynchronous accumulator. A performance table is provided 
in their paper for comparison between their construction and other asymmetric accumulator types.

\begin{table*}[h]
%\renewcommand{\arraystretch}{1.3}
% \extrarowheight 
\caption{Known Cryptographic Accumulator Implementations %{\color{red}TO DO: Add URLs for each source code product.}
}
\label{tab:knownApp}
%\centering
% Some packages, such as MDW tools, offer better commands for making tables
% than the plain LaTeX2e tabular which is used here.
\resizebox{\linewidth}{!}{%
\begin{tabular}%{.5\linewidth}
{|l|l|}
\hline
\textbf{Asymmetric} & \textbf{Symmetric} \\ \hline
RSA Based Accumulator Source Code \cite{RSA-BM-code} & Bloom Filter Source Code \cite{Bloom-Code} \\ \hline
Bilinear Map Based Accumulator Source Code \cite{RSA-BM-code} & Cuckoo Filter Source Code \cite{cuckoo-filter-github} \\ \hline
Merkle Tree Based Accumulator Source Code \cite{Bitcoin-Merkle-Code,IAIK-Merkle-Code} & %{\color{red} Add Braavos, and t-SDH} 
\\ \hline
\end{tabular}
}
%\vspace*{-.50cm}
\end{table*}

Table~\ref{tab:knownApp} provides a list of known implementations of cryptographic accumulators in C/C++. In the RSA and Bilinear-Map implementations, three libraries were used: FLINT \cite{flint-library} %{\color{red}\cite{CITEME}}
to facilitate modular arithmetic over large integers, Crypto++ \cite{cryptopp-library} %{\color{red} \cite{CITEME}} 
for thread-safe cryptographic operations, and the DCLXVI library \cite{dclxvi-library} %{\color{red} \cite{CITEME}} 
for the elliptic curve computations required for the bilinear map accumulator \cite{tremel2013real}. 
%It should be noted that 

Note that an RSA-based accumulator is only collision free when accumulating prime numbers. A prime representative generator is needed to support the accumulation of sets containing composite values \cite{tremel2013real}. %{\color{red} \cite{CITEME}}. 
In the RSA implementation provided, the random oracle prime representative generator suggested by  Bari\'{c}
%Barić 
and Pfitzmann in \cite{baric1997collision} was used \cite{tremel2013real}. Both Merkle Tree implementations use SHA-256 hashing to calculate the Merkle root \cite{IAIK-Merkle-Code}. %{\color{red} \cite{CITEME}}. 
The Cuckoo filter implementation \cite{Cuckoo-Code} uses the OpenSSL software library to perform MD5 and SHA-1 hashing \cite{openssl-library}. 

\section{\uppercase{Conclusion}}\label{sec:conclusion}
We provided a concise guide 
to cryptographic accumulators. 
We presented  their security and so-called  optional properties.
Also, we discussed the effects of different optional properties on accumulator performance in terms of space, time, and communication complexity.

Cryptocurrencies were early adopters of cryptographic accumulators. For instance, Bitcoin uses a Bloom Filter 
for set-membership testing of transactions \cite{bitcoin-bloom} 
. Bitcoin also uses the
Merkle Block \cite{bitcoin-merkleBlock} to confirm the validity of transactions 
without having to download and/or maintain the entire copy of blockchain by every participating node.
In Zerocoin \cite{miers2013zerocoin}, 
a CL-RSA-B based accumulator is implemented to represent a set of all transactions that have been committed to blockchain \cite{miers2013zerocoin}. 
The accumulator additionally eliminates trackable linkage of addresses and facilitates anonymous and untraceable public transactions \cite{miers2013zerocoin}. 

Other applications of cryptographic accumulators include maintaining a certificate revocation list,  blacklisting or whitelisting user credentials in an authentication system,
and generating important client groups such as lists of high risk and/or bankrupt clients.

Cryptographic accumulators can also help maintain dataset  privacy    while sharing. They can be implemented to share and maintain a list of such users without actually revealing their identities. 
This property could simplify sharing of critical data sharing with third parties. Cryptographic accumulators can also be used in offline ID verification, credit score checking, and medical-data verification.

\section*{\uppercase{Acknowledgments}}
This material is based upon work supported by, or in part by,the National Science Foundation  (NSF) under grants 
CCF-1562659, CCF-1562306, CCF-1617690, CCF-1822191, CCF-1821431, and The Scientific and Technological Research Council of Turkey (TUBITAK). The views and conclusions contained herein are those of the authors and should not be interpreted as  representing the official policies, views, or endorsements, either expressed or implied, of the
NSF or TUBITAK.

Significant support from the University of Tennessee at Chattanooga's Center for Excellence in Applied Computational Science and Engineering (SimCenter) is also gratefully acknowledged.

\bibliography{ozcelik-et-al-2021-v1}

\bibliographystyle{apalike}

\end{document}